\documentclass[FBSedit,FBSmath,ecsub]{FBSsuppl}
\usepackage{amsfonts}
\usepackage{amssymb}
\usepackage{epsfig}

\title{\hfill{\small{\bf MKPH-T-02-13}}\\
The Two-Nucleon System beyond the Pion Threshold}
\author{M. Schwamb\thanks{\textit{E-mail address:} 
 schwamb@kph.uni-mainz.de} and  H. Arenh\"ovel}
\institute{Institut f\"ur Kernphysik, Johannes Gutenberg-Universit\"at
Mainz, D-55099 Mainz, Germany} 
\begin{document}

\maketitle
\vspace*{-0.5cm}
\begin{abstract}
A model is developed for the hadronic and electromagnetic interaction
 in the two-nucleon system above pion threshold in the framework of meson,
 nucleon and $\Delta$ degrees of freedom. It is based on time-ordered
 perturbation theory and  includes full meson
 retardation in potentials and exchange currents
  as well as loop contributions to the nucleonic one-body current. 
 Results for some selected processes are discussed.
\end{abstract}

\section{Introduction}
 At present, a very interesting topic in the field of medium energy physics
is devoted to the role of effective degrees of freedom (d.o.f.)
 in hadronic systems in  terms of nucleon, meson and isobar d.o.f.
 and their connection to the underlying quark-gluon dynamics of
 QCD.  For the study of 
this basic question, the two-nucleon system provides an 
important test laboratory, because it is  the simplest 
system for the study of the nucleon-nucleon interaction,
  the role of medium effects due to two-body 
 operators and the relevance of offshell effects, i.e.\ the change of single
 particle properties in the nuclear medium.
Moreover,  the deuteron is of specific relevance as an effective neutron
target. Within the present effective description, one is able  to obtain
  a reasonable, though not perfect description of the possible 
 hadronic and electromagnetic reactions in the two-nucleon sector
below $\pi$ threshold.
However, for energies beyond  $\pi$ thres\-hold up to the $\Delta$ region
additional complications occur. Present
 state-of-the-art  models
 should incorporate not only a realistic NN interaction, but also
 a dynamical treatment of the $\Delta$ isobar. Moreover, in view of
 the requirement of
 unitarity, the various possible  reactions  should be described
 within one consistent framework,
 because they are linked by the optical theorem.
 In the past, we have started to realize
 this ambitious project  within a {\it retarded}
 coupled channel $NN-N\Delta$ approach, based on a three-body scattering
 approach with nucleon, $\Delta$ and meson degrees of freedom
\cite{ScA98a,ScA98b,ScA01a,ScA01b,ScA01c}.

\section{The Model}
For energies up to the $\Delta$ region, 
 our model Hilbert space ${\cal H}^{[2]}$ 
  consists  of three orthogonal subspaces
${\cal H}^{[2]} = {\cal H}^{[2]}_{\bar N} \oplus
{\cal H}^{[2]}_{\Delta} \oplus 
{\cal H}^{[2]}_{X}\, $, 
where ${\cal H}^{[2]}_{\bar N}$ contains two (bare) nucleons, 
${\cal H}^{[2]}_{\Delta}$ one nucleon and one $\Delta$ resonance, 
and ${\cal H}^{[2]}_{X}$ two nucleons and one meson
$X \in \{ \pi, \rho, \sigma, \delta, \omega, \eta\}$.

Concerning the hadronic part,  the basic  interactions in our model 
 are $XNN$  and $\pi N \Delta$ vertices.
   Inserting these into the Lippmann-Schwinger equation,
 one obtains after some straightforward algebra \cite{ScA01a} effective
 hadronic  interactions acting in 
 ${\cal H}^{[2]}_{\bar N} \oplus  {\cal H}^{[2]}_{\Delta}$ which contain,
 apart from pion-nucleon loop contributions to the nucleonic self energy, 
 {\it retarded} one-boson exchange (OBE) mechanisms describing the transitions
 $NN \rightarrow NN$, $NN \rightarrow N\Delta$ and  
 $N\Delta \rightarrow N\Delta$. 
  Due to the  nonhermiticity and  nonlocality of a retarded operator, which
 becomes moreover complex  above pion  threshold,  in most 
 applications the  simple static approximation
 is used by neglecting the energy transfer
  by the exchanged meson. The corresponding static operators
are much easier to handle, but  one encounters on the other hand
 at least two serious problems. Above
 pion-threshold, unitarity is violated due to the absence of  
 singularities in the static 
 operators.  Moreover, in the past it turned out 
 that even the simplest photonuclear reaction, namely
 deuteron photodisintegration, cannot be  described even
 qualitatively within a consistent
 static framework \cite{TaO89,WiA93}.

In  our explicit realization, we use for the
 retarded $NN$ interaction
  the Elster potential \cite{ElF87} which takes into account
 in  addition one-pion loop
 diagrams  in order to fulfil unitarity above pion threshold. Therefore,
 one has to distinguish between bare and physical nucleons
 (see \cite{ScA01a} for details). Concerning the
 transitions $NN \rightarrow N\Delta$ and 
 $N\Delta \rightarrow N\Delta$, we take besides retarded
 pion exchange   static $\rho$ exchange into account. Moreover,
 the interaction of two nucleons in the deuteron channel in the presence of
 a spectator pion (the  $\pi d$ channel) is also considered.
 This mechanism is necessarry for satisfying unitarity, because otherwise
no asymptotic free $\pi d$ state would exist. By a
 suitable box renormalization \cite{GrS82}, we are able to obtain approximate
 phase equivalence between the Elster potential and our coupled
 channel approach below pion threshold.
 
Similarly, the basic electromagnetic interactions  consist of
 baryonic and mesonic one-body currents as well as  vertex and 
 Kroll-Rudermann contributions \cite{ScA01b,ScA01c}. These currents are,
 together with the $\pi NN$ vertex, the basic building blocks
 of the corresponding effective
 current operators. The latter contain beside the ordinary spin-,
 convection- and spin-orbit currents full retarded pionic 
 meson exchange currents
 and electromagnetic loop contributions, where the latter
 can be interpreted as  off-shell contributions  to the baryonic
 one-body current \cite{ScA01c}. Moreover, static $\rho$ MEC as well as
 $\Delta$ MEC contributions are taken into
 account. It can be shown \cite{ScA01c} that concerning the pionic part
 gauge invariance is fulfiled  in leading order of $1/M_N$.
 This is a consequence of the fact that the hadronic pion-nucleon loop
contributions, the retarded $\pi$-exchange contribution to the NN interaction,
 the electromagnetic loop contributions, and the $\pi$ MEC are based on the
same $\pi NN$ vertex. 

\begin{figure}[b]
\centerline{\epsfig{file=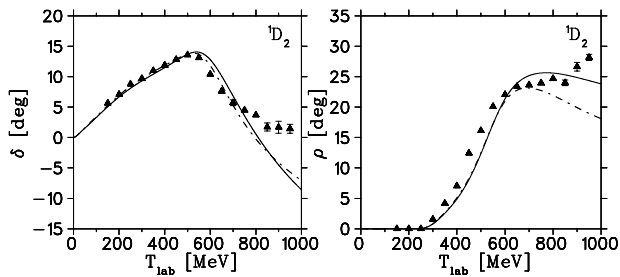,width=230pt}}
\caption[]{Phase shift $\delta$ and inelasticity $\rho$ for the
 $^1D_2$ $NN$-channel in comparison with experiment (solution SM97 of
 Arndt {\it et al.}~{\protect\cite{Arn98}})
for two  potential models: dash-dotted curve: static approach, based
 on the Bonn-OBEPR potential {\protect\cite{MaH87}},
 full curve: retarded approach. See {\protect\cite{ScA01a}} for further
 details.}
\label{nnphase}
\end{figure}

\section{Results}
The hadronic $\Delta$ parameters and the M1 $\gamma N \Delta$ coupling
 are simultaneously fitted  to the $M_{1+}(3/2)$
 multipole of pion  photoproduction on the nucleon,  to the $P_{33}$ channel
 of pion-nucleon scattering and to the $^1D_2$ channel of  nucleon-nucleon
 scattering \cite{ScA01a,ScA01b}. In Fig.\ \ref{nnphase}, our results for the
 $^1D_2$ phase shift and inelasticity are depicted. 
 We obtain a good description at least up to  about $T_{lab}=800$ MeV.
 This is of particular relevance for deuteron photodisintegration where
 the most important contribution to the unpolarized cross section
 in the $\Delta$ region  comes
from  the $^1D_2$ channel.
 Concerning the other partial $NN$ waves, the overall description
 is only fairly well
 \cite{ScA01a}. Therefore, we have started  to construct from scratch 
 an  interaction 
 model whose parameters are fitted to the phase shifts and inelasticities
 of {\it all} relevant $NN$  partial waves for  energies
 up to about 1 GeV. However, as of yet  we have not found a parameter set
 which would lead to a considerable improvement. One reason for this failure
  may be the fact that our effective baryon-baryon model
is basically a one-boson exchange model which does not take into account
  more complicated
mechanisms like the crossed two-pion-exchange. Moroever,
the $\pi N$ interaction has to be improved.
 Forthcoming studies on that topic are essential, especially for clarifying
the role of final state interactions (FSI) in hadronic and electromagnetic
breakup reactions on the deuteron for energies beyond the $\pi$ threshold.

\begin{figure}[t]
\centerline{\epsfig{file=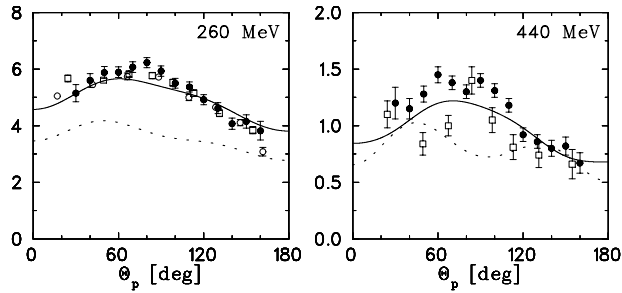,width=230pt}}
\caption[]{Differential cross section of deuteron photodisintegration
 for two photon energies $k_{lab}$ as  function
 of the c.m.\ proton angle $\theta_p$: dotted curve: result of
 Wilhelm {\it et al.} in static approach {\protect \cite{WiA93}},
 full curve: retarded approach of {\protect \cite{ScA01b}}.
 Offshell contributions to the nucleonic one-body current are
 included, too  {\protect \cite{ScA01c}}.
Experimental data from {\protect \cite{CrA96}} ($\bullet$), 
 {\protect \cite{ArG84}} (open box) and
 {\protect \cite{BlB95}} ($\circ$).}
\label{fig_gd}
\end{figure}

 As next, we discuss very briefly deuteron photodisintegration.
  The starting point of our consideration is the
   static approach of Wilhelm {\it et al.} \cite{WiA93} 
 which is based on the Bonn-OBEPR potential \cite{MaH87}.
 Similar to our  present approach, there is 
 no free parameter in the calculation of the 
  photodisintegration process in \cite{WiA93}. As is evident
 form Fig.\ \ref{fig_gd}, Wilhelm {\it et al.} 
 clearly fail in describing the  data. One obtains
  a considerable underestimation of the cross section in the $\Delta$ peak.
 Moreover, a dip structure around $90^{\circ}$ occurs at higher energies which
is not present in the data. This feature is also present in other static
approaches like the unitary three-body model of Tanabe and Otha
\cite{TaO89}. 
 On the other hand,  these problems
 in the differential cross section  
 vanish almost completely in a retarded
 approach \cite{ScA01b, ScA01c}.  However, some discrepancies in  polarization
 observables like the linear photon asymmetry $\Sigma$ or the polarization
 $P_y(p)$ of the outgoing proton
  are still present \cite{ScA01b, ScA01c}. It is known that polarization
observables may be  sensitive to small reaction mechanisms which are
suppressed in the unpolarized cross section. Therefore, we assume that   
the failure
of our model  in describing $\Sigma$ and $P_y(p)$ may be attributed to
the present treatment of the hadronic interaction which needs to be 
improved as has been outlined above.

A  very interesting topic  is the exploration of the spin asymmetry 
 of the total cross section of the nucleon and light nuclei  which
  determines the GDH-sum rule \cite{Ger65,DrH66}
 \begin{equation}\label{gdh}
I^{GDH}= \int dk \frac{\sigma^P(k) - \sigma^A(k)}{k} 
 = 4 \pi \kappa^2 \frac{e^2}{M^2}S \,\, 
\end{equation}
 and  
which is at present under investigation  experimentally  \cite{AhA01}.
In (\ref{gdh}), $\sigma^{P/A}(k)$ denotes the total photoabsorption
cross section on a target   with mass $M$, spin $S$ and anomalous magnetic
moment $\kappa$ with the photon spin parallel (antiparallel) to the target
spin. 

 Due to the lack of a free neutron target, a measurement  of the GDH-sum rule
 on the deuteron is of specific significance  because it has been suggested in
the past that one can measure the spin asymmetry of the total photoabsorption
 cross section of the neutron by using light nuclei like the deuteron. 
This would rest on the assumptions that (i)  the polarized deuteron
constitutes an effective polarized neutron target and (ii)
one has to suppose that the spin asymmetry on the deuteron  is dominated by
the quasi free pion-production, so that binding and FSI effects
  from the presence of the spectator nucleon can be neglected,
 resulting in an incoherent sum of proton and neutron 
contributions. 

However,  both assumptions
 are questionable.
 First of all, due to the existence of the $D$ state component
in the deuteron wave function, even in a completely polarized deuteron target,
the neutron is not  completely polarized. With respect to the second 
assumption, one has to recall that, 
  in contrast to the nucleon  case,  for the GDH-integral on the
{\it deuteron} also the energy region {\it below} $\pi$ threshold has to be
taken into account due to the existence of the breakup 
channel $\gamma d \rightarrow NN$ 
which is absent in the photoabsorption  on a free nucleon.
 As has been pointed
out in \cite{ArK97},
 there exists a strong anticorrelation between pion production
and photodisintegration, because the latter reaction yields a very
large {\it negative} contribution to the GDH-integral in (\ref{gdh}) 
  near the breakup threshold. This facts explains at least partially the
 large difference between the GDH-value on the deuteron 
 ($I^{GDH}_d=0.65$ $\mu$b) and the sum of the GDH-value on proton 
 and neutron yielding 438 $\mu$b. 
 Morover, the effect of final
state interactions in $\pi$ production on the deuteron
  cannot at all be neglected  \cite{DaA02}
so that a simple spectator model picture is not appropriate for
the evaluation of the spin asymmetry on the deuteron.

\begin{figure}[t]
\centerline{\epsfig{file=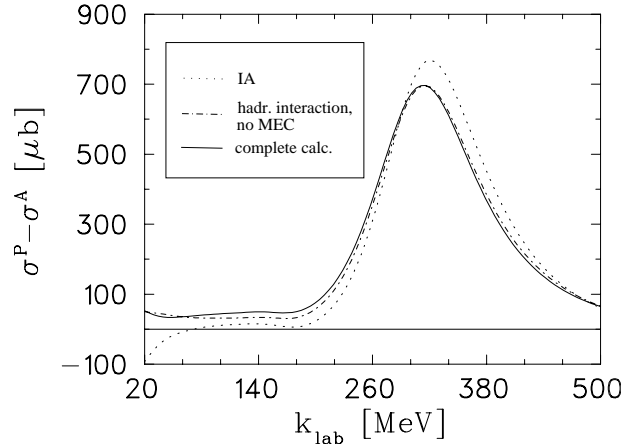,width=230pt}}
\caption[]{Preliminary result for the spin asymmetry on the deuteron
within our retarded approach as a function of the photon energy $k_{lab}$
in the laboratory frame. The dotted curve shows the result within a simple
spectator approach. In the dashed-dotted curve, hadronic backscattering 
mechanisms are added. In the full curve, additional two-body meson exchange
currents are taken into account. A part of the Born term contributions
in pion production has been neglected.}
\label{gdh_fig}
\end{figure}

Due to these facts, it is obvious that an extraction of the neutron
amplitude from the deuteron reaction cannot be performed in a model
independent manner. Nevertheless, the experimental investigation
 of the GDH-sum rule on the deuteron is very important
 because the measurement of the spin asymmetry
 is for itself of great interest since it will give us much more
detailed information on the underlying dynamics
 than just the unpolarized cross section.  Therefore, it 
 may serve in the future
 as a stringent test for existing models (see for example
 Fig. \ref{gdh_fig}) in the two-nucleon sector.

\section{Outlook}
 In the future, we plan  to apply the present model to other
  reactions, especially 
 electrodisintegration.  Conceptually, 
 we have to  improve our hadronic interaction model. Moreover, additional
 d.o.f. like the Roper, the $D_{13}$ and the $S_{11}$  resonances should 
 be taken into account if one wants to consider higher energies.

\end{document}